\documentclass[sigconf]{acmart}

\usepackage{booktabs}
\usepackage{amsmath}
\usepackage{footmisc}
\usepackage{mathrsfs}
\usepackage{graphicx}
\usepackage{dblfloatfix}
\usepackage{multirow}
\usepackage{algorithm}
\usepackage{algpseudocode}

\usepackage{multicol}
\usepackage{enumitem}
\usepackage{subcaption}
\usepackage{tabularx}
\usepackage{array}
\usepackage{float}
\usepackage{balance}
\usepackage{hyperref}

\AtBeginDocument{%
  \providecommand\BibTeX{{%
    \normalfont B\kern-0.5em{\scshape i\kern-0.25em b}\kern-0.8em\TeX}}}

\copyrightyear{2026}
\acmYear{2026}
\setcopyright{cc}
\setcctype{by-nc-nd}
\acmConference[KDD '26]{Proceedings of the 32nd ACM SIGKDD Conference on Knowledge Discovery and Data Mining V.2}{August 09--13, 2026}{Jeju Island, Republic of Korea}
\acmBooktitle{Proceedings of the 32nd ACM SIGKDD Conference on Knowledge Discovery and Data Mining V.2 (KDD '26), August 09--13, 2026, Jeju Island, Republic of Korea}
\acmDOI{10.1145/3770855.3818206}
\acmISBN{979-8-4007-2259-2/2026/08}
\begin{document}
\title{Hierarchical Residual Policy Optimization for Generative Recommendations}

\author{Kaifeng Guo}
\authornote{This work was done during Kaifeng Guo's internship at Kuaishou Technology.}
\affiliation{%
  \institution{City University of Hong Kong}
  \city{Hong Kong}
  \country{China}}
\email{kaifenguo2-c@my.cityu.edu.hk}

\author{Yiming Yang}
\affiliation{%
  \institution{Kuaishou Technology}
  \city{Beijing}
  \country{China}}
  \email{yangyiming03@kuaishou.com}

\author{Jingtong Gao}
\affiliation{%
  \institution{City University of Hong Kong}
  \city{Hong Kong}
  \country{China}}
\email{jt.g@my.cityu.edu.hk}

\author{Guolei Zeng}
\affiliation{%
  \institution{Independent}
  \city{Fuzhou}
  \country{China}}
\email{guolei.zeng@gmail.com}

\author{Fukang Yang}
\affiliation{%
  \institution{Kuaishou Technology}
  \city{Beijing}
  \country{China}}
\email{yangfukang03@kuaishou.com}

\author{Yukang Liang}
\affiliation{%
  \institution{Kuaishou Technology}
  \city{Beijing}
  \country{China}}
\email{liangyukang@kuaishou.com}

\author{Peng Jiang}
\orcid{0009-0000-7636-6453}
\affiliation{%
  \institution{Kuaishou Technology}
  \city{Beijing}
  \country{China}}
\email{jiangpeng11@kuaishou.com}

\author{Qingpeng Cai}
\authornote{Corresponding authors.}
\affiliation{%
  \institution{Kuaishou Technology}
  \city{Beijing}
  \country{China}}
\email{caiqingpeng@kuaishou.com}

\author{Xiangyu Zhao}
\authornotemark[2]
\affiliation{%
  \institution{City University of Hong Kong}
  \city{Hong Kong}
  \country{China}}
\email{xianzhao@cityu.edu.hk}

\renewcommand{\shortauthors}{Kaifeng Guo et al.}

\begin{abstract}

Generative recommenders select items by autoregressively decoding semantic identifiers (SIDs), whose token positions induce a coarse-to-fine hierarchy over the item space. 
In practice, SID decoders are trained via supervised next-token prediction, which imitates logged trajectories rather than directly optimizing downstream utility. 
This motivates post-training with outcome feedback to guide decoding toward higher utility. 
However, logged feedback is only observed for the final exposed item, causing most post-training methods to operate at the item level and broadcast the same terminal signal across all SID tokens. 
As a result, token-level credit assignment becomes sparse, high-variance, and layer-dependent.
To this end, we propose \textbf{H}ierarchical \textbf{R}esidual \textbf{P}olicy \textbf{O}ptimization (HRPO), a post-training framework that converts item-level outcomes into dense, token-aligned learning signals for conservative token-wise improvement. 
Specifically, HRPO first estimates SID prefix-level utilities via group-wise reward smoothing over feature-based user clusters. It then decomposes these utilities into residual token credits and accumulates them into credit-to-go signals. 
Finally, \textbf{R}esidual-\textbf{R}eturn \textbf{P}olicy \textbf{O}ptimization (RRPO) optimizes the residual credits using clipped updates, group-normalized advantages, and KL regularization to preserve stability. 
Experiments on a public dataset and an online A/B test in a large-scale commercial system show consistent gains in session-level utility and key business metrics. 
Source code and the archived artifact are available for reproduction~\footnote{Code: \url{https://github.com/Applied-Machine-Learning-Lab/KDD2026-HRPO}; DOI: \url{https://doi.org/10.5281/zenodo.20269678}.}.

\end{abstract}

\keywords{Generative Recommendation, Hierarchical Autoregressive Modeling, Token-level Credit Assignment, Preference Alignment, Recommender System Simulation}

\ccsdesc[500]{Information systems~Recommender systems}

\maketitle

\section{Introduction}

Generative recommendation has emerged as an alternative to direct item selection by generating item identifiers autoregressively, token by token~\cite{tiger,onerec,minionerec}. 
Many approaches adopt Semantic Identifiers (SIDs), which assign each item a short sequence of discrete codes derived from a structured scheme such as a tree or trie~\cite{tdm}. 
Under this formulation, token positions induce a coarse-to-fine hierarchy over the item space: early tokens (SID-prefix) select a coarse partition, while later tokens progressively refine within that partition until the sequence uniquely identifies an item. 
This hierarchical factorization decomposes an enormous item universe into a small number of sequential token decisions, substantially improving scalability~\cite{minionerec}.

In practice, SID generators are typically trained in two stages~\cite{onerec,minionerec}. 
First, supervised fine-tuning (SFT) learns a valid SID decoder by maximizing next-token likelihood on logged SID trajectories, yielding a stable and readily deployable initialization~\cite{tiger,minionerec}. 
However, SFT is inherently likelihood-based: it models the conditional distribution of the next token given a logged prefix under the behavior policy, rather than directly optimizing the outcome-based utility used in downstream evaluation. 
In contrast, large-scale recommender and advertising systems are evaluated using multi-objective outcomes (e.g., engagement, conversion, and cost-related signals) that are observed after an item is generated and exposed, reflecting implicit user preferences and business constraints~\cite{jannach2023mors,mcmahan2013adclick,bottou-jmlr-2013}. 
This mismatch between imitation-driven training and outcome-driven evaluation motivates a second-stage post-training procedure, which initializes from the SFT model and leverages logged outcomes to improve expected utility~\cite{onerec,swaminathan2015crm}.

\begin{figure}[t]
  \centering
  \includegraphics[width=\columnwidth]{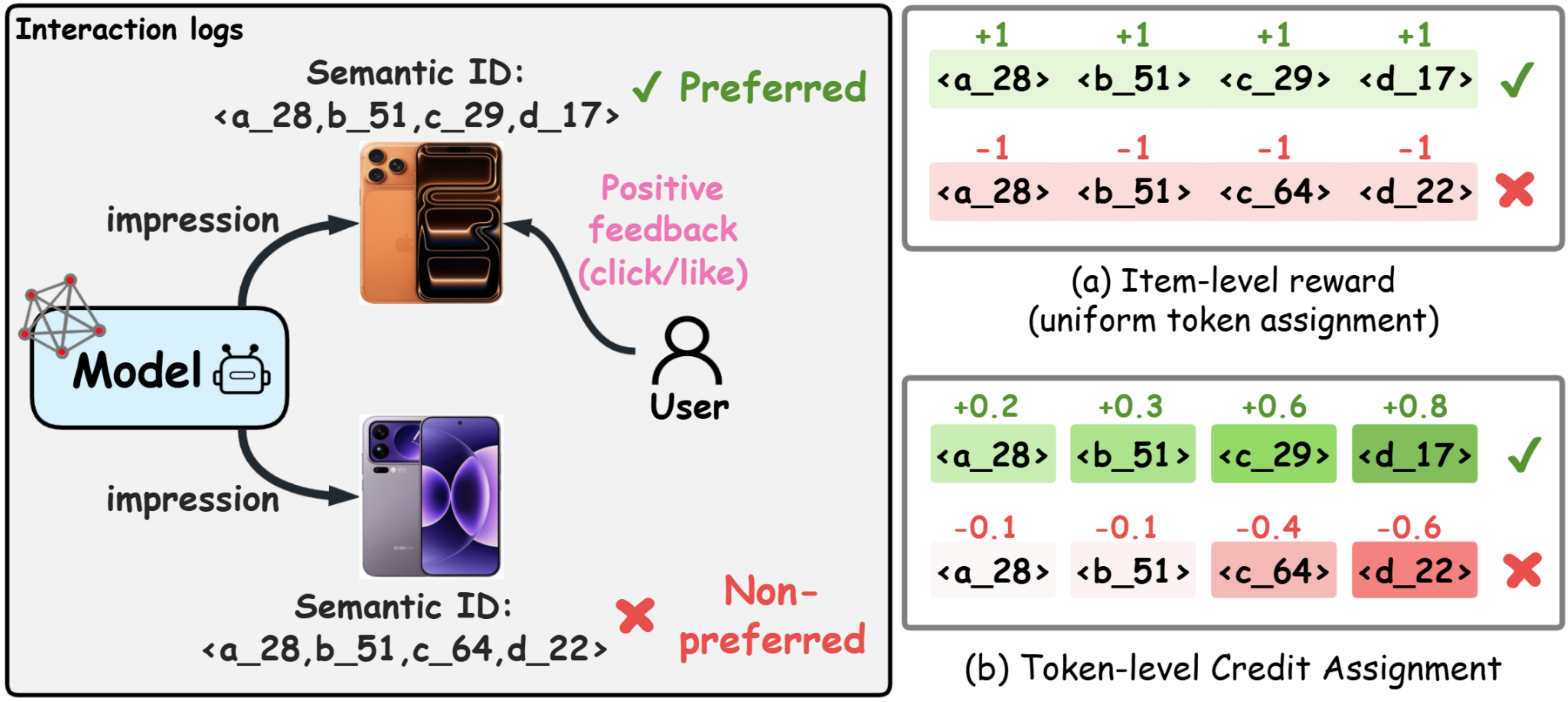}
  \vspace{-2.0em}
  \caption{Token-level credit assignment for hierarchical Semantic IDs (SIDs).
  Logged feedback is only observed after a complete SID path is generated.
  (a) Uniformly assigning the terminal item reward to every token yields undifferentiated supervision.
  (b) Token-level credit assignment decomposes the outcome into layer-specific credits, enabling targeted post-training along the SID hierarchy.}
  \vspace{-1.5em}
  \label{fig:token_credit_assignment}
\end{figure}

A central challenge is how to transform item-level outcomes into effective learning signals for token-by-token SID decoding. 
In existing SID-based generators, post-training often remains item-centric, mapping the terminal outcome of the exposed item to token-level updates~\cite{onerec,minionerec,sdpo,sprec,rere}. 
Common approaches include (i) uniformly broadcasting the same terminal reward to all tokens along the decoded SID path, and (ii) PPO-style optimization, in which token advantages are derived from terminal feedback~\cite{williams1992simple,ranzato2016sequence,bahdanau2017actorcritic,ppo}. 
Figure~\ref{fig:token_credit_assignment} illustrates why such item-level treatments can be suboptimal for hierarchical SID decoding. 
Because many items share early SID prefixes and diverge only at deeper positions (Figure~\ref{fig:token_credit_assignment}, left), uniform broadcasting yields weak and non-discriminative supervision (Figure~\ref{fig:token_credit_assignment}a) and may even induce conflicting gradients on shared prefixes. 
More generally, item-level updates obscure which SID layer should be adjusted to improve utility. 
This motivates token-level credit assignment (Figure~\ref{fig:token_credit_assignment}b), which decomposes item outcomes into layer-specific credits, enabling post-training to update decisions at the appropriate depth of the SID hierarchy.

While token-level credit assignment is desirable (Figure~\ref{fig:token_credit_assignment}b), deriving reliable token-wise learning signals from exposure-limited logs is challenging. 
A practical signal should be estimable for long-tail SID prefixes, attributable despite delayed end-of-sequence outcomes, and comparable across candidates at the same token position to support stable optimization. 
However, logged recommendation data rarely satisfies all three requirements simultaneously. 
We summarize three key obstacles.
\textbf{(1) SID-prefix sparsity.} Interactions are highly uneven across SID prefixes; for long-tail prefixes, observations are insufficient to reliably estimate prefix-level utility, resulting in noisy and unstable credits~\cite{steck2011popularity,abdollahpouri2019popbias}.
\textbf{(2) Delayed attribution.} Feedback is observed only after a SID path is completed (i.e., for the finally exposed item), so terminal outcomes do not directly indicate which intermediate token decisions should be adjusted~\cite{sutton2018rl,williams1992simple,ranzato2016sequence,bahdanau2017actorcritic}.
\textbf{(3) Non-comparable scales.} Credits conditioned on different prefixes can differ substantially in scale and variance, making position-wise advantages difficult to compare and destabilizing PPO-style updates in large-scale settings~\cite{schulman2016gae,ppo}.

To address these challenges, we propose \textbf{HRPO} (Hierarchical Residual Policy Optimization), a post-training framework that transforms item-level outcomes into dense, token-aligned learning signals for conservative token-wise improvement. 
HRPO integrates three key ideas. 
First, it employs feature-based grouping to obtain smoothed utility estimates for SID prefixes under limited log coverage. 
Second, it introduces residual token credits to enable incremental, layer-aware attribution. 
Third, it constructs position-aligned credit-to-go signals to stabilize optimization~\cite{ng1999shaping}. 
Building on these token-aligned signals, we instantiate \textbf{RRPO} as HRPO’s conservative optimization objective. 
RRPO performs clipped updates with group-normalized advantages and KL regularization, targeting stable improvement in large-scale systems~\cite{ppo}.

Finally, we benchmark HRPO offline on KuaiRand~\cite{kuairand}, an open public dataset, with session-level evaluation. 
We further report results from a post-launch online A/B test conducted in a large-scale advertising system, demonstrating consistent improvements under real-world operational constraints~\cite{kohavi2012trustworthy}.

Our main contributions are:
\begin{itemize}[leftmargin=*]
    \item We identify a token-level credit assignment bottleneck in post-training SID generators when learning from item-level outcomes. In particular, reward broadcasting provides weak supervision on shared prefixes, while prefix-conditioned credits are often incomparable across positions, destabilizing PPO-style updates.
    \item We propose \textbf{HRPO}, a structured post-training framework that densifies sparse feedback via feature-based grouping and derives token-aligned learning signals through residual credits and credit-to-go returns. We further instantiate \textbf{RRPO} as a conservative optimization objective.
    \item We validate HRPO on a public dataset and complementary online A/B test evidence, demonstrating both controlled gains in simulation and positive production effects on target business metrics.
\end{itemize}

\section{Preliminaries}
\label{sec:prelim}

This section formalizes the problem setting and introduces the notation used throughout the paper.
We highlight the key design choices and assumptions needed for offline learning from exposure-limited logs.
We then summarize SID decoding as the underlying sequential decision process that our method builds on for post-training.
\subsection{Task Formulation and Logged Multi-Feedback}

We consider generative recommendation~\cite{tiger,onerec,minionerec}, in which a policy selects
an item by generating a structured identifier sequence conditioned
on a request context. Each training instance corresponds to one
exposure in offline logs~\cite{li2011unbiased,swaminathan2015crm}.

Let $\mathcal{U}$ be the user set, $\mathcal{I}$ the item set, and
$\mathcal{X}$ the space of request contexts. For each impression, we
represent the record as $(u_i, x_i, v_i, \mathbf{b}_i)$ with $u_i \in \mathcal{U}$, context
$x_i \in \mathcal{X}$ (user/profile features, truncated interaction
history, and situational features), displayed item $v_i \in
\mathcal{I}$, and multi-feedback vector $\mathbf{b}_i \in \{0,1\}^M$, where $M$ is the number of feedback
channels and $b_{i,m}=1$ indicates that the $m$-th feedback signal is recorded
(e.g., click and long-view). The
offline dataset is
\[
\mathcal{D} = \{(u_i, x_i, v_i, \mathbf{b}_i)\}_{i=1}^N.
\]

To optimize multi-objective behavior~\cite{jannach2023mors}, we scalarize $\mathbf{b}_i$ via a
configurable linear utility
\begin{equation}
  r(\mathbf{b}_i) = \sum_{m=1}^{M} w_m b_{i,m},
  \label{eq:scalar_utility}
\end{equation}
where $\{w_m\}_{m=1}^M \subset \mathbb{R}$ are pre-specified weights. For
conciseness, we define $r_i := r(\mathbf{b}_i)$.

Our goal is to learn a generative policy $\pi_\theta$ that maximizes
the expected utility of exposed items under deployment constraints.

\begin{figure*}[t]
  \centering
  \includegraphics[
    width=0.75\textwidth,
    trim=0 0 0 0,
    clip
  ]{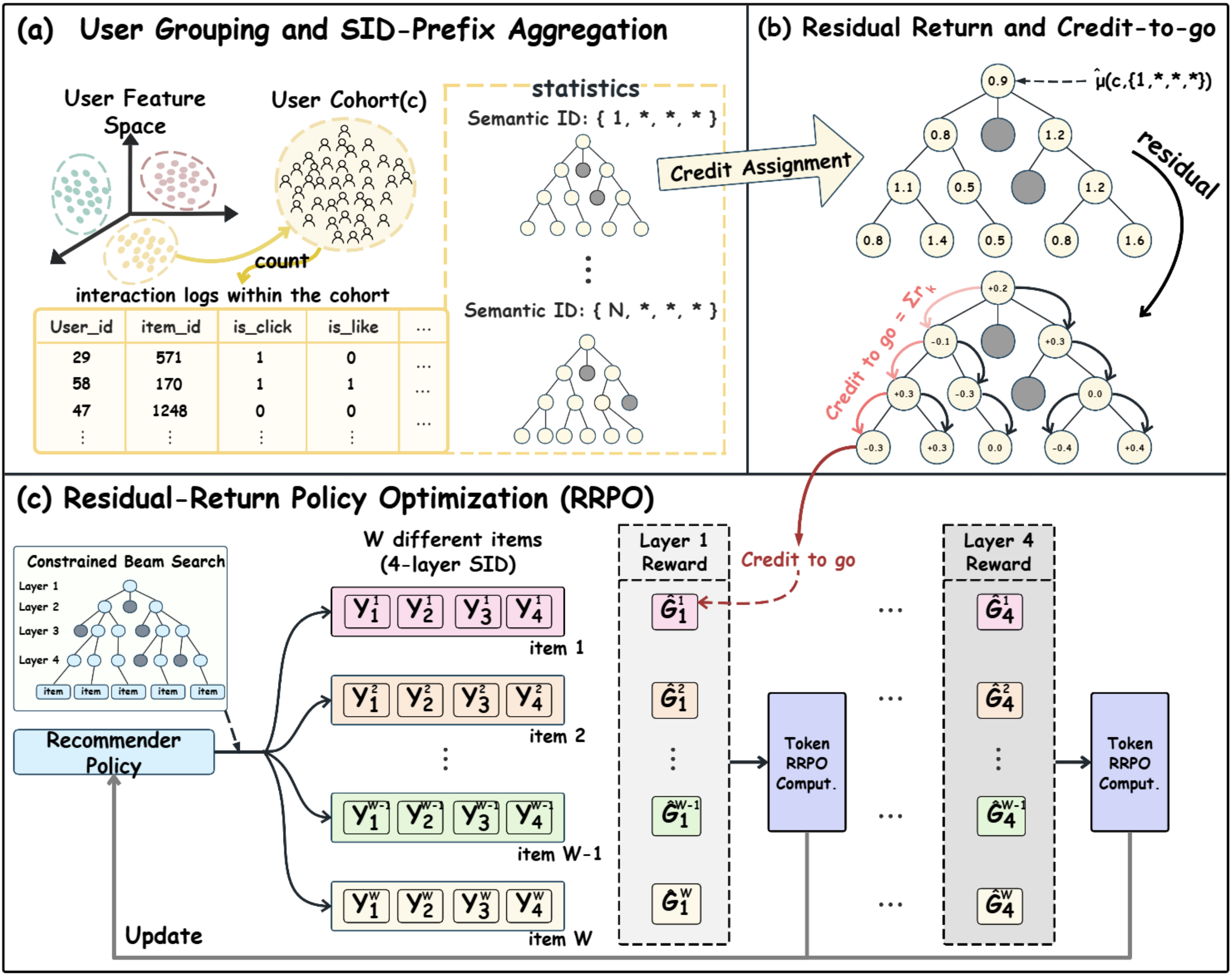}
  \vspace{-0.4em}
\caption{Overview of HRPO.}
  \vspace{-0.6em}
  \label{fig:cohort_residual}
\end{figure*}

\subsection{Semantic Identifiers (SID)}
Following recent work on SID-based generative retrieval~\cite{tiger,onerec,minionerec},
each item $v\in\mathcal{I}$ is mapped to a fixed-length SID,
i.e., a depth-$L$ token path produced by an encoding function $\phi$:
\begin{equation}
\phi(v)=\mathbf{y}=(y_1,\ldots,y_L)\in \mathcal{V}_1\times\cdots\times \mathcal{V}_L,
\label{eq:sid-map}
\end{equation}
where $\mathcal{V}_t$ denotes the token vocabulary at depth $t$.

For any $t\in\{0,1,\ldots,L\}$, we call the first $t$ tokens of an SID a SID prefix (or prefix),
denoted by $\mathbf{y}_{1:t}$:
\begin{equation}
\mathbf{y}_{1:t}\triangleq (y_1,\ldots,y_t), \qquad \mathbf{y}_{1:0}\triangleq \epsilon,
\label{eq:sid-prefix}
\end{equation}
where $\epsilon$ denotes the empty sequence.

A prefix $\mathbf{y}_{1:t}$ corresponds to a node in the SID trie and represents the subset of items
whose SIDs share the same prefix:
\begin{equation}
\mathcal{I}(\mathbf{y}_{1:t}) \triangleq \{\, v\in\mathcal{I}\mid \phi(v)_{1:t}=\mathbf{y}_{1:t}\,\}.
\label{eq:sid-subtree}
\end{equation}
Intuitively, decoding proceeds by selecting tokens to refine the SID-prefix, thereby progressively shrinking $\mathcal{I}(\mathbf{y}_{1:t})$ until a unique item is reached at depth $L$. In the logged data, each exposed item $v_i$ can be equivalently represented by its SID path $\mathbf{y}_i=\phi(v_i)$, yielding records $(u_i, x_i, \mathbf{y}_i, r_i)$ as needed.

\subsection{Constrained Generation}
Each item is represented by a semantic identifier (SID) $\mathbf{y}=(y_1,\ldots,y_L)$.
We generate $\mathbf{y}$ autoregressively:
\begin{equation}
\pi_\theta(\mathbf{y}\mid x)=\prod_{t=1}^L \pi_\theta(y_t\mid x, \mathbf{y}_{1:t-1}),
\label{eq:sid_ar}
\end{equation}
where $y_{<t}$ denotes the SID prefix (the first $t\!-\!1$ tokens).
Decoding is validity-constrained by the SID trie: at step $t$, we restrict the next token to an admissible set
$\mathcal{A}(y_{<t})$ induced by the current SID-prefix and renormalize the distribution over this set.

\section{Methodology}
\label{sec:method}
In this section, we introduce the HRPO framework and describe its main components step by step.

\subsection{Overview}
\label{sec:method_overview}

Fig.~\ref{fig:cohort_residual} summarizes \textsc{HRPO}, a post-training framework for
SID-based generative recommendation. The key difficulty is that offline logs provide
utility feedback only for the final exposed item (a full SID path), while the
policy makes decisions at every SID token. \textsc{HRPO} addresses this action--feedback
mismatch by turning logged item-level outcomes into dense, position-aligned
token-level learning signals, and by updating the policy conservatively to mitigate
offline distribution shift~\cite{li2011unbiased,swaminathan2015crm}.

Given logged interactions and a deployable base generator (e.g., an SFT model),
\textsc{HRPO} proceeds in three stages, aligned with Fig.~\ref{fig:cohort_residual}:
\textbf{(a) User grouping and SID-prefix aggregation.}
We deterministically map each user to a feature-defined group (cohort) $c$ and aggregate
logged outcomes on the SID trie to estimate a smoothed expected utility for each prefix,
denoted $\hat\mu(c,\mathbf{y}_{1:t})$.
This step transforms sparse, heavy-tailed prefix statistics into stable utilities that
can be queried for any prefix encountered during decoding.
\textbf{(b) Residual return and credit-to-go.}
For any candidate SID path $\mathbf{y}=(y_1,\ldots,y_L)$, we convert prefix utilities
into token-level residual credits (marginal gains from choosing $y_t$ given the
current prefix) and accumulate them into a position-aligned credit-to-go
$\hat G_t(c,\mathbf{y})$, which represents the remaining utility beyond the current
prefix baseline. This yields comparable token-wise signals across different candidates
at the same position.
\textbf{(c) Residual-Return Policy Optimization (RRPO).}
For each logged request context $x$, we construct a small proposal set
$Y(x)=\{\mathbf{y}^{(w)}\}_{w=1}^{W}$ of valid SID paths using a frozen
reference policy $\pi_{\theta_{\mathrm{old}}}$, optionally including the
logged SID path and adding unique trie-valid candidates decoded under
$\pi_{\theta_{\mathrm{old}}}$. Importantly, $\pi_{\theta_{\mathrm{old}}}$ is
an internal reference/proposal policy for RRPO, rather than the unknown
behavior policy that produced the historical log. We then update the policy
with RRPO, a PPO-style clipped token objective that uses within-set,
position-wise normalized advantages derived from $\hat{G}_t$, while
regularizing toward the same frozen reference policy to prevent
over-optimistic updates on out-of-distribution tokens.

\subsection{User Grouping and SID-prefix Aggregation}
\label{sec:method_grouping}

Offline logs provide feedback only for the exposed final item, while token-level learning requires utility estimates for intermediate SID prefixes. A naive approach estimates a prefix utility by averaging terminal utilities over all log records that share the prefix, but this is unreliable for long-tail prefixes~\cite{steck2011popularity,abdollahpouri2019popbias,efron1975stein}. To better align terminal outcomes with the sequential token decisions made during trie-constrained decoding, we estimate smoothed prefix utilities that can be queried for any encountered prefix (Figure~\ref{fig:cohort_residual}(a)).

The challenge is that prefix statistics are highly imbalanced and heavy-tailed: many long-tail prefixes (and many
user--prefix pairs) have only a few exposures, making per-user estimates noisy or undefined.
To obtain well-defined, low-variance prefix utilities while retaining meaningful personalization, we pool data within
feature-defined user cohorts.
Since this aggregation is used in deployment, we use a deterministic grouping function; 
concretely, we assign each user $u$ to a cohort $c(u)\in\{0,1,\dots\}$ based on a small set of stable user features (e.g., activity level, spend tier, or region):
\begin{equation}
c(u)=g(u),
\label{eq:cohort_def}
\end{equation}
where $g(\cdot)$ is a fixed mapping implemented in the system; Appendix~\ref{app:cohort_sensitivity} reports the concrete KuaiRand features, cohort support, and grouping sensitivity.

For each record with cohort $c=c(u)$ and path $\mathbf{y}$, we update cohort-conditional prefix statistics along the
trie path:
\begin{equation}
n(c,y_{1:t}) \leftarrow n(c,y_{1:t})+1,\qquad
s(c,y_{1:t}) \leftarrow s(c,y_{1:t})+r,\quad t\in[1,L],
\label{eq:prefix_stats}
\end{equation}

Empirical means $s(c,y_{1:t})/n(c,y_{1:t})$ can be unstable in the long tail, so we apply a lightweight shrinkage with
pseudo-count $\alpha$ toward the global mean $\mu_{\mathrm{glob}}$:
\begin{equation}
\hat{\mu}(c,y_{1:t})=\frac{s(c,y_{1:t})+\alpha\,\mu_{\mathrm{glob}}}{n(c,y_{1:t})+\alpha},\qquad
\mu_{\mathrm{glob}}=\frac{1}{|\mathcal{D}|}\sum_{(u,x,v,\mathbf{b})\in\mathcal{D}} r(\mathbf{b}).
\label{eq:shrink}
\end{equation}

In implementation, for rarely observed cohort-specific prefixes, we default to a broader (global) estimate. Finally, we score a candidate SID path $\mathbf{y}$ under context $x$ (user $u$) using the terminal-prefix estimate
$\hat{\mu}\big(c(u),y_{1:L}\big)$.

\subsection{Residual Return and Credit-to-go}
\label{sec:method_residual_return}

Given cohort-conditioned prefix utilities on the SID trie (Section~\ref{sec:method_grouping}), we construct position-aligned residual returns for SID tokens and accumulate them into a credit-to-go signal. These dense, token-level targets address the action--feedback mismatch in log-only training and enable stable within-proposal normalization in RRPO (Figure~\ref{fig:cohort_residual}(b)).

Offline feedback is only observed at the terminal leaf (the exposed final item),
whereas the policy acts at every token position during trie-constrained decoding.
A common workaround in existing approaches---including terminal-reward RL and
item-level reranking based post-training---is to treat the whole SID path as a
single action and broadcast the same terminal reward to all tokens on the path~\cite{williams1992simple,ranzato2016sequence,ouyang2022instructgpt,ppo}.
However, this uniform assignment blurs which SID layer (or prefix transition)
actually contributed to the outcome, and it often yields noisy, high-variance
token updates because all positions share an undifferentiated supervision signal.

Our solution is to residualize prefix utilities along the prefix sequence
(Figure~\ref{fig:cohort_residual}(b)).
Specifically, we view $\hat\mu(c, y_{1:t})$ as the estimated value of the trie node
corresponding to prefix $y_{1:t}$ under cohort $c$, and assign each token $y_t$ the
marginal utility gain of moving from the parent prefix $y_{1:t-1}$ to the child prefix
$y_{1:t}$:
\begin{equation}
\begin{aligned}
\hat r_t(c,\mathbf{y}) &= \hat\mu(c,y_{1:t})-\hat\mu(c,y_{1:t-1}),\\
y_{1:0} &\triangleq \epsilon,\qquad \hat\mu(c,\epsilon)\triangleq \mu_{\mathrm{glob}}.
\end{aligned}
\label{eq:residual_credit}
\end{equation}
This residual credit removes the cohort-specific prefix baseline and attributes
utility locally to the token choice at position $t$.
Moreover, it is additive and consistent with trie traversal: the sum telescopes to
\(\sum_{t=1}^{L}\hat r_t(c,\mathbf{y})=\hat\mu(c,y_{1:L})-\mu_{\mathrm{glob}}\), i.e., token credits form a
decomposition of the terminal prefix score up to a constant baseline.

Using only the instantaneous increment $\hat r_t$ is insufficient: two candidates may have similar local
increments at position $t$ but very different remaining utility depending on how the suffix will unfold.
Conversely, broadcasting the terminal score to every position discards the current prefix state and
degenerates to undifferentiated supervision. We therefore aggregate residual credits from the current
position onward, which preserves the prefix baseline at $y_{1:t-1}$ while accounting for the remaining
potential beyond it.

Specifically, we define the cumulative residual return (credit-to-go) as:
\begin{equation}
\hat G_t(c,\mathbf{y})=\sum_{k=t}^{L}\hat r_k(c,\mathbf{y}),
\label{eq:residual_return_def}
\end{equation}
which telescopes to
\begin{equation}
\hat G_t(c,\mathbf{y})=\hat\mu(c,y_{1:L})-\hat\mu(c,y_{1:t-1}).
\label{eq:residual_return_telescoping}
\end{equation}
Eq.~\eqref{eq:residual_return_telescoping} shows that $\hat G_t$ measures the remaining utility beyond the current prefix baseline $\hat\mu(c,y_{1:t-1})$.
As a result, $\hat G_t$ is naturally comparable across candidates at the same position $t$ (same $t$, different $\mathbf{y}$) while remaining consistent with the cohort-conditioned terminal prefix score. This makes $\hat G_t$ a suitable learning signal for within-group normalization and conservative token-wise optimization.

\subsection{Residual-Return Policy Optimization (RRPO)}
\label{sec:method_rrpo}

RRPO performs conservative offline updates of the trie-constrained generative policy by treating the
token-level credit-to-go $\hat G_t(c,\mathbf{y})$ (Section~\ref{sec:method_residual_return}) as the return signal.
Offline policy refinement is sensitive to distribution shift, and under trie constraints, increasing the
probability of an early token can redirect decoding to a different feasible subtree, inducing amplified
changes in later-token distributions~\cite{tdm,tiger,schulman2015trpo,ppo}. RRPO mitigates these risks with (i) within-proposal, position-wise
normalization, (ii) PPO-style clipped updates, and (iii) an explicit trust-region penalty.

For a fixed context $x$ (user $u$), let $c\triangleq c(u)$.
We first sample a proposal set $Y(x)=\{ \mathbf{y}^{(w)} \}_{w=1}^{W}$ from $\pi_{\theta_{\mathrm{old}}}$, where
$W$ is the proposal size.
For each candidate, define $\hat G_t^{(w)} \triangleq \hat G_t(c,\mathbf{y}^{(w)})$
(Eq.~\eqref{eq:residual_return_def}).
We normalize $\hat G_t^{(w)}$ within the proposal group at each position $t$ to obtain a scale-free advantage:
\begin{equation}
A_t^{(w)}
=
\frac{
\hat G_t^{(w)}-\frac{1}{W}\sum_{j=1}^{W}\hat G_t^{(j)}
}{
\mathrm{Std}\!\left(\{\hat G_t^{(j)}\}_{j=1}^{W}\right)+\varepsilon
}.
\label{eq:token_group_adv}
\end{equation}

We then form token-level likelihood ratios under the same autoregressive factorization and trie-valid
conditional distributions as in Eq.~\eqref{eq:sid_ar}.
For candidate $w$ at position $t$, with prefix $\mathbf{y}^{(w)}_{1:t-1}$ and token $y_t^{(w)}$, the ratio is
\begin{equation}
\rho_{w,t}(\theta)=
\exp\Big(
\log \pi_\theta(y^{(w)}_t \mid x, \mathbf{y}^{(w)}_{1:t-1})
-\log \pi_{\theta_{\mathrm{old}}}(y^{(w)}_t \mid x, \mathbf{y}^{(w)}_{1:t-1})
\Big).
\label{eq:token_ratio}
\end{equation}
To prevent overly large offline updates, we apply PPO-style clipping:
\begin{equation}
\bar\rho_{w,t}(\theta) \;=\;
\mathrm{clip}\!\big(\rho_{w,t}(\theta),\,1-\epsilon_{\text{clip}},\,1+\epsilon_{\text{clip}}\big).
\label{eq:token_ratio_rrpo}
\end{equation}

RRPO maximizes a clipped token-wise surrogate weighted by $A_t^{(w)}$, and penalizes deviation from
$\pi_{\theta_{\mathrm{old}}}$ via a trust-region KL term:
\begin{equation}
\begin{aligned}
\max_{\theta}\quad
&\frac{1}{W}\sum_{w=1}^{W}\sum_{t=1}^{L}
\min\!\Big(\rho_{w,t}(\theta)\,A_t^{(w)},\ \bar\rho_{w,t}(\theta)\,A_t^{(w)}\Big)
-\beta_{\mathrm{KL}}\,\mathrm{KL}_{\mathrm{TR}}(\theta),
\end{aligned}
\label{eq:rrpo_obj}
\end{equation}
\begin{equation}
\mathrm{KL}_{\mathrm{TR}}(\theta)
=\mathbb{E}_{\mathbf{y}\sim \pi_{\theta_{\mathrm{old}}}(\cdot\mid x)}
\Big[\log \pi_{\theta_{\mathrm{old}}}(\mathbf{y}\mid x)-\log \pi_\theta(\mathbf{y}\mid x)\Big].
\label{eq:tr_kl}
\end{equation}

Training alternates between (i) proposal generation under $\pi_{\theta_{\mathrm{old}}}$,
(ii) computing $\hat G_t$ and $A_t$ from cohort-conditioned prefix utilities
(Eqs.~\eqref{eq:shrink}--\eqref{eq:token_group_adv}),
and (iii) updating $\pi_\theta$ by maximizing Eq.~\eqref{eq:rrpo_obj}.
We periodically synchronize $\theta_{\mathrm{old}} \leftarrow \theta$; full pseudocode is provided in
Appendix~\ref{app:hrpo_algorithms} for reference.

\section{Experiments}
\label{sec:exp}

In this section, we conduct a comprehensive evaluation of \textsc{HRPO} for generative recommendation. To thoroughly assess \textsc{HRPO}'s effectiveness and understand when and why it works, we aim to answer the following research questions:
\begin{itemize}[leftmargin=*]
     \item \textbf{RQ1:} How does \textsc{HRPO} compare to mainstream post-training methods for optimizing immediate (single-shot) utility versus long-horizon (session-level) return?
    \item \textbf{RQ2:} How sensitive are the results to key hyperparameters and design choices?
    \item \textbf{RQ3:} How do individual components of \textsc{HRPO} contribute to the final performance?
    \item \textbf{RQ4:} Do the improvements persist online with validated gains?
\end{itemize}

\subsection{Dataset and Simulator Details}
\label{sec:exp_dataset_sim}

We construct offline training and evaluation logs from the public KuaiRand dataset~\cite{kuairand}.
We use the \textit{Pure} subset and sessionize the impression stream into user-day sessions for sequential recommendation.
Each impression record contains a user $u$, context $x$, an exposed item $v$, and multi-behavior feedback $\mathbf{b}$, which is scalarized into a utility
$r(\mathbf{b})$ via Eq.~(\ref{eq:scalar_utility}).
Table~\ref{tab:kuairand_summary} summarizes the resulting sessionized logs used throughout our experiments.

Our objective is to better align token-level SID decisions with users' latent preferences under multi-objective feedback.
In practice, preference signals are revealed through sequential interactions, and platform utility is typically a scalarization of multiple behaviors
(Eq.~(\ref{eq:scalar_utility})) rather than a single immediate metric.
Purely offline evaluation from static logs is limited for diagnosing such interaction-induced effects and for comparing methods that require interaction.
Therefore, following prior work, we additionally use a controlled simulator built on top of the same public logs to provide step-wise feedback and state transitions,
enabling session-level evaluation of cumulative simulator reward and simulator-interactive baselines.
In the public KuaiSim rollout tables, we report the simulator's click-based Avg./Total reward as the primary reward metric, and report Long separately as an auxiliary behavior rate.
We treat the simulator as a proxy rather than a perfect replica of production, and corroborate conclusions with validated online A/B evidence in production. Accordingly, simulator results are used for controlled relative comparison under a shared environment, whereas production conclusions are drawn from the online A/B test.

We use the public KuaiRand-Pure logs and KuaiSim rollout protocol following the simulator's standard preprocessing and evaluation setup.

\begin{table}[t]
  \caption{Summary statistics of the sessionized KuaiRand-Pure logs used in our experiments.}
  \label{tab:kuairand_summary}
    \vspace{-1.0em}
  \centering
  \small
  \setlength{\tabcolsep}{3pt}
  \renewcommand{\arraystretch}{0.86}
  \begin{tabular}{l r}
    \toprule
    \textbf{Statistic} & \textbf{Value} \\
    \midrule
    Time span & 2022-04-09 $\sim$ 2022-05-08 (30 days) \\
    \#Users & 19{,}574 \\
    \#Items (videos) & 5{,}659 \\
    \#Interactions & 1{,}341{,}250 \\
    \#Sessions (user-day) & 303{,}472 \\
    Avg.\ interactions/user & 68.5 (median 50) \\
    Avg.\ interactions/session & 4.42 (median 2, p90 10) \\
    Avg.\ active days/user & 15.5 (median 15) \\
    Density $|\mathcal{D}|/(|\mathcal{U}||\mathcal{I}|)$ & 0.012109 \\
    \bottomrule
  \end{tabular}
\end{table}

\subsection{Implementation Details}
\label{sec:exp_setup}

We implement all methods in PyTorch and evaluate them in KuaiSim~\cite{kuaisim}.
For SID-generating policies, the action is a fixed-depth SID path with depth $L{=}4$ and per-level vocabulary size $|\mathcal{V}_t|{=}32$.
Decoding follows the shared SID trie admissible set $y_t \in \mathcal{A}(y_{<t})$ (Eq.~\eqref{eq:sid_ar}); by default, we use constrained beam search (beam width 50) and take the top-1 beam.

All SID post-training baselines and \textsc{HRPO} share the same small Transformer decoder backbone
(3 blocks, hidden width $d{=}128$, 4 attention heads, maximum history length 50) and start from the same
SFT checkpoint trained with teacher-forced next-token prediction for 5 epochs.
For log-only post-training, all methods use AdamW with learning rate $1{\times}10^{-5}$, weight decay $0.01$,
batch size 1024, gradient clipping at 1.0, and one epoch over the offline logs.
For \textsc{HRPO}, we use proposal size $W{=}18$, PPO clip $\epsilon_{\mathrm{clip}}{=}0.2$,
KL coefficient $\beta_{\mathrm{KL}}{=}0.1$, smoothing pseudo-count $\alpha{=}100$,
and synchronize the frozen reference policy every 20 update steps.
Experiments are run on a single NVIDIA RTX~4090 GPU (24GB), representing a small-model regime in our offline study.
For production-backbone scaling diagnostics, we additionally report 0.005B--0.2B model families in Appendix~\ref{app:scaling_diagnostics}.

\subsection{Baselines}
\label{sec:exp_baselines}

We benchmark \textsc{HRPO} against representative baselines spanning three families:
(i) score-based sequential recommenders trained on offline logs,
(ii) SID-generating policies optimized by offline post-training objectives, and
(iii) simulator-interactive RL methods trained in KuaiSim.
All methods are evaluated under the same experimental pipeline, with the shared backbone, decoding, and optimization settings summarized in Sec.~\ref{sec:exp_setup}.

\begin{itemize}[leftmargin=*]
    \item \textbf{Score-based (offline logs):} DT (behavior cloning)~\cite{chen2021decisiontransformer}, GRU4Rec~\cite{gru4rec}, SASRec~\cite{sasrec}, TIGER~\cite{tiger}.
    \item \textbf{SID post-training (offline logs):} SFT, DPO~\cite{dpo}, S-DPO~\cite{sdpo}, SPRec~\cite{sprec}, GRPO~\cite{rere}.
    \item \textbf{Simulator-interactive RL (KuaiSim):} TD3~\cite{td3}, DDPG~\cite{ddpg}, A2C~\cite{mnih2016async}, HAC~\cite{hac}, and an interactive DT~\cite{chen2021decisiontransformer}.
\end{itemize}

Across all methods, we follow KuaiSim's whole-session evaluation protocol.
EpisodeLen corresponds to KuaiSim's \emph{Depth}, i.e., the number of request--feedback interactions before the user leaves the session (averaged over evaluation sessions).
In the public KuaiSim rollout evaluation, we use the simulator's click-based immediate reward, denoted by $r_t=b_t^{\mathrm{click}}\in\{0,1\}$, as the primary reward metric for Avg./Total reward.
We report the undiscounted session return (Total reward) as the average sum of immediate click rewards per session, $\frac{1}{N}\sum_{i=1}^N\sum_{t=1}^{T_i} r_{i,t}$, and Avg.\ reward as the per-request mean return, i.e., Total reward divided by EpisodeLen.
We also report Coverage as the number of distinct exposed items, and impression-level Click/Long rates as the empirical frequencies of the corresponding binary behavior indicators, respectively.
Long is reported as an auxiliary behavior metric rather than being included in the Avg./Total reward columns.
For one-step evaluation, EpisodeLen $=1$ and Avg.\ reward equals Total reward.

\subsection{Overall Performance (RQ1)}
\label{sec:exp_results}

Tables~\ref{tab:offline_wholesession}--\ref{tab:interactive_wholesession} summarize three complementary evaluation settings:
(i) offline training with session-level rollouts, (ii) offline training with one-step evaluation (EpisodeLen $=1$), and
(iii) simulator-interactive training with session-level rollouts.

Table~\ref{tab:offline_wholesession} reports whole-session performance for log-only methods.
(1) Score-based recommenders improve over Random but remain behind SID-generating policies, suggesting that validity-constrained SID decoding provides a stronger action space for long-horizon optimization in this simulator.
(2) Among SID-generating post-training baselines, \textsc{HRPO} achieves the best session return:
Total reward increases from 8.137 (\textsc{SPRec}) to 10.528 and EpisodeLen increases from 14.14 to 15.33.
Coverage stays comparable to other offline baselines (56\% vs.\ 57\% for \textsc{SPRec}),
indicating that the gain is not explained by overly concentrating exposure on a small set of items.

To isolate immediate utility, Table~\ref{tab:onestep_offline} evaluates a single decision (EpisodeLen $=1$).
\textsc{HRPO} attains the highest one-step reward (0.548 vs.\ 0.462 for \textsc{GRPO}),
consistent with stronger local preference alignment at the first decision.
The lower one-step Coverage indicates that \textsc{HRPO} is more selective in assigning probability mass to high-utility items, and over longer horizons these local gains can compound across multiple steps.

Table~\ref{tab:interactive_wholesession} compares \textsc{HRPO} to simulator-interactive KuaiSim RL baselines under a matched interaction budget.
These baselines achieve moderate returns but exhibit very low Coverage (e.g., 3\%--12\%),
which is consistent with mode-seeking behavior under nearest-neighbor action realization.
\textsc{HRPO} achieves both the best return (Total reward 10.746) and substantially higher Coverage (57\%),
indicating more stable exploration--exploitation behavior.

\begin{table}[t]
\caption{Offline training with session-level rollout evaluation on KuaiSim. Bold indicates the best value; underlined indicates the second-best value.}
  \label{tab:offline_wholesession}
    \vspace{-1.0em}
  \centering
  \small
  \setlength{\tabcolsep}{3.5pt}
  \renewcommand{\arraystretch}{1.12}
  \resizebox{\linewidth}{!}{%
  \begin{tabular}{lcccccc}
    \hline
    \textbf{Method} & \textbf{EpisodeLen} & \textbf{Avg.\ reward} & \textbf{Total reward} & \textbf{Coverage} & \textbf{Click} & \textbf{Long} \\
    \hline
    Random & 11.16 & 0.1898 & 2.1170 & \textbf{100\%} & 18.97\% & 12.32\% \\
    DT & 12.79 & 0.4274 & 5.4660 & \underline{91\%} & 42.74\% & 37.19\% \\
    GRU4Rec & 11.52 & 0.2573 & 2.9650 & 84\% & 25.73\% & 21.79\% \\
    SASRec & 12.09 & 0.3351 & 4.0530 & 72\% & 33.51\% & 26.75\% \\
    TIGER & 13.85 & 0.5442 & 7.5370 & 70\% & 54.42\% & 46.78\% \\
    SFT & 14.02 & 0.5642 & 7.9110 & 60\% & 56.41\% & 49.80\% \\
    DPO & 13.99 & 0.5610 & 7.8490 & 58\% & 56.10\% & 49.49\% \\
    S-DPO & 13.64 & 0.5240 & 7.1470 & 60\% & 52.39\% & 45.52\% \\
    SPRec & \underline{14.14} & \underline{0.5753} & \underline{8.1370} & 57\% & \underline{57.52\%} & \underline{51.29\%} \\
    GRPO & 14.07 & 0.5690 & 8.0060 & 58\% & 56.90\% & 50.49\% \\
    \textbf{HRPO} & \textbf{15.33} & \textbf{0.6869} & \textbf{10.5280} & 56\% & \textbf{68.69\%} & \textbf{63.04\%} \\
    \hline
  \end{tabular}}

\end{table}

\begin{table}[t]
  \caption{Offline training with one-step evaluation on KuaiSim (EpisodeLen $=1$), isolating immediate utility. Bold indicates the best value; underlined indicates the second-best value.}
  \label{tab:onestep_offline}
    \vspace{-1.0em}
  \centering
  \small
  \setlength{\tabcolsep}{3.5pt}
  \renewcommand{\arraystretch}{1.12}
  \resizebox{\linewidth}{!}{%
  \begin{tabular}{lcccccc}
    \hline
    \textbf{Method} & \textbf{EpisodeLen} & \textbf{Avg.\ reward} & \textbf{Total reward} & \textbf{Coverage} & \textbf{Click} & \textbf{Long} \\
    \hline
    Random & 1.0 & 0.2130 & 0.2130 & \textbf{100\%} & 21.30\% & 13.00\% \\
    SFT & 1.0 & 0.4510 & 0.4510 & \underline{28\%} & 45.10\% & 34.20\% \\
    S-DPO & 1.0 & 0.4600 & 0.4600 & 23\% & 46.00\% & 36.80\% \\
    SPRec & 1.0 & 0.4570 & 0.4570 & 21\% & 45.70\% & 35.30\% \\
    GRPO & 1.0 & \underline{0.4620} & \underline{0.4620} & 23\% & \underline{46.20\%} & \underline{36.90\%} \\
    \textbf{HRPO} & 1.0 & \textbf{0.5480} & \textbf{0.5480} & 15\% & \textbf{54.80\%} & \textbf{44.70\%} \\
    \hline
  \end{tabular}}

\end{table}

\begin{table}[t]
  \caption{Simulator-interactive training with session-level evaluation on KuaiSim. Bold indicates the best value; underlined indicates the second-best value.}
  \label{tab:interactive_wholesession}
    \vspace{-1.0em}
  \centering
  \small
  \setlength{\tabcolsep}{3.5pt}
  \renewcommand{\arraystretch}{1.12}
  \resizebox{\linewidth}{!}{%
  \begin{tabular}{lcccccc}
    \hline
    \textbf{Method} & \textbf{EpisodeLen} & \textbf{Avg.\ reward} & \textbf{Total reward} & \textbf{Coverage} & \textbf{Click} & \textbf{Long} \\
    \hline
    Random & 11.16 & 0.1898 & 2.1170 & \textbf{100\%} & 18.97\% & 12.32\% \\
    TD3 & 13.52 & 0.5128 & 6.9330 & 10\% & 51.27\% & 45.41\% \\
    A2C & 12.19 & 0.3468 & 4.2270 & 50\% & 34.68\% & 27.60\% \\
    DDPG & 13.71 & 0.5359 & 7.3470 & 3\% & 53.59\% & 48.98\% \\
    HAC & 13.97 & 0.5604 & 7.8270 & 3\% & 56.04\% & \underline{50.84\%} \\
    DT & \underline{13.98} & \underline{0.5624} & \underline{7.8630} & 12\% & \underline{56.24\%} & 50.17\% \\
    \textbf{HRPO} & \textbf{15.50} & \textbf{0.6931} & \textbf{10.7460} & \underline{57\%} & \textbf{69.31\%} & \textbf{63.79\%} \\
    \hline
  \end{tabular}}

\end{table}

\subsection{Hyperparameter Sensitivity (RQ2)}
\label{sec:exp_sensitivity}

\begin{figure}[t]
  \centering
  \begin{subfigure}[t]{0.48\linewidth}
    \centering
    \includegraphics[width=\linewidth]{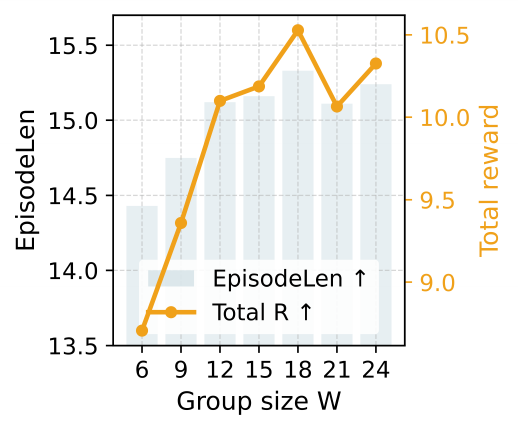}
    \vspace{-2.0em}
    \caption{Group size $W$.}
    \label{fig:group_size}
  \end{subfigure}
  \hfill
  \begin{subfigure}[t]{0.48\linewidth}
    \centering
    \includegraphics[width=\linewidth]{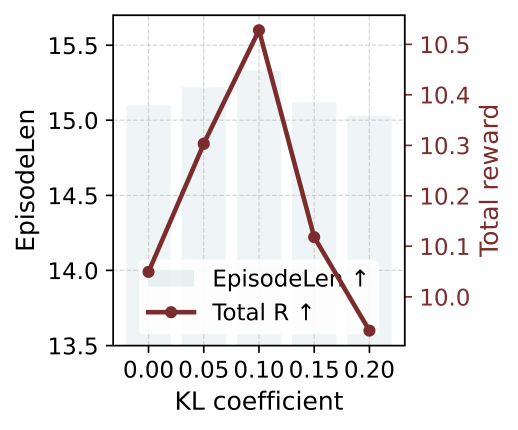}
    \vspace{-2.0em}
    \caption{KL $\beta_{\mathrm{KL}}$.}
    \label{fig:kl_coeff}
  \end{subfigure}

  \vspace{0.6em}

  \begin{subfigure}[t]{0.48\linewidth}
    \centering
    \includegraphics[width=\linewidth]{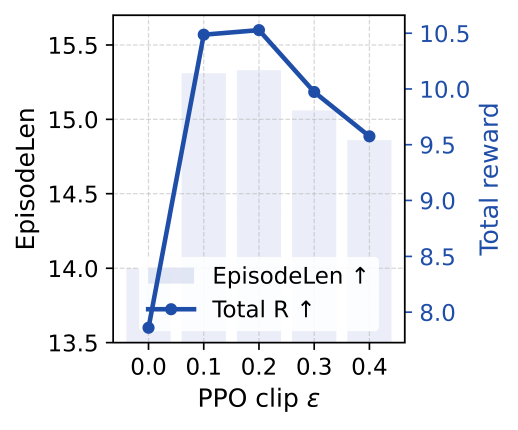}
    \vspace{-2.0em}
    \caption{Clip $\epsilon_{\mathrm{clip}}$.}
    \label{fig:clip_eps}
  \end{subfigure}
  \hfill
  \begin{subfigure}[t]{0.48\linewidth}
    \centering
    \includegraphics[width=\linewidth]{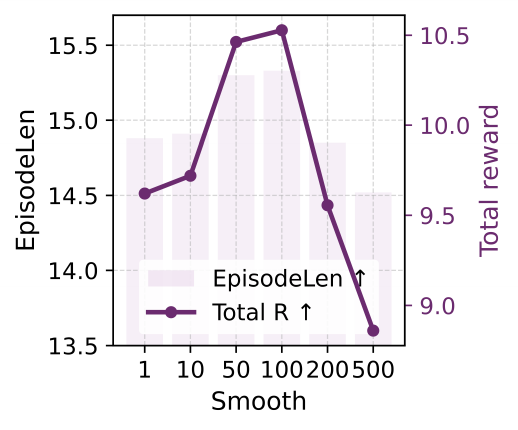}
    \vspace{-2.0em}
    \caption{Smooth $\alpha$.}
    \label{fig:smooth}
  \end{subfigure}

  \caption{Hyperparameter sensitivity under offline training with session-level evaluation.
  Bars indicate EpisodeLen; the line indicates Total reward.}
  \label{fig:hparam_sensitivity}
\end{figure}

Figure~\ref{fig:hparam_sensitivity} summarizes sensitivity to four key knobs in HRPO/RRPO under offline training with session-level evaluation.
Across sweeps (varying one hyperparameter at a time; others fixed to the default settings summarized in Sec.~\ref{sec:exp_setup}), performance is stable in a broad neighborhood, and the best region consistently corresponds to moderate aggregation/regularization that balances robustness and learning capacity.

For group-based updates (Figure~\ref{fig:hparam_sensitivity}A), increasing group size $W$ improves return and EpisodeLen up to a saturation region (best around $W\!\approx\!18$ in our setup).
Larger groups strengthen within-context normalization and expose more informative hard negatives, making token credits more contrastive; beyond saturation, extra candidates add diminishing information and can introduce overly hard/redundant negatives, increasing gradient noise and slightly hurting stability.

For KL regularization (Figure~\ref{fig:hparam_sensitivity}B), a non-zero $\beta_{\mathrm{KL}}$ is beneficial (peak around $\beta_{\mathrm{KL}}\!\approx\!0.10$), indicating that constraining updates to stay close to the frozen reference/SFT policy prevents overfitting to a narrow set of hard negatives under exposure-limited logs.
However, overly large $\beta_{\mathrm{KL}}$ makes the update too conservative and caps improvement in both EpisodeLen and return.

For PPO clipping (Figure~\ref{fig:hparam_sensitivity}C), moderate $\epsilon_{\mathrm{clip}}$ yields the best trade-off.
Too small $\epsilon_{\mathrm{clip}}$ suppresses likelihood-ratio changes and slows learning, while too large $\epsilon_{\mathrm{clip}}$ allows abrupt policy shifts that amplify token-wise variance across SID layers and can destabilize decoding.
This aligns with RRPO's goal of conservative, incremental improvement: clipping controls step size when token credits are high-variance, especially for early SID tokens that gate large subtree changes.

For shrinkage in prefix-utility estimation (Figure~\ref{fig:hparam_sensitivity}D), $\alpha$ controls a variance--bias trade-off in the long tail.
Weak shrinkage leaves rare-prefix utilities noisy and propagates high-variance credits, whereas overly strong shrinkage collapses prefix differences toward the global mean and reduces the discriminativeness needed for layer-specific credit assignment; the best region corresponds to sufficient smoothing to denoise without washing out prefix-level structure.

Taken together, these sweeps support that HRPO is not brittle: it benefits from (i) sufficiently informative groups ($W$), (ii) conservative policy improvement (KL and clipping), and (iii) calibrated smoothing for long-tail prefixes ($\alpha$), aligning with the roles of group normalization, conservative optimization, and robust reward estimation in our design.

\begin{table}[t]
  \caption{HRPO ablations (compact). Total reward is the undiscounted episode return; the one-step column reports Avg.\ reward with EpisodeLen $=1$.}
  \label{tab:ablation_hrpo_AB_compact}
  \vspace{-1.0em}
  \centering
  \small
  \setlength{\tabcolsep}{3.5pt}
  \renewcommand{\arraystretch}{1.05}
  \resizebox{1\linewidth}{!}{
  \begin{tabular}{lccc c}
    \toprule
    & \multicolumn{3}{c}{\textbf{Session-level}} & \textbf{One-step} \\
        \cmidrule(lr){2-4}\cmidrule(lr){5-5}
        \textbf{Variant} & \textbf{EpisodeLen} & \textbf{Avg.\ reward} & \textbf{Total reward} & \textbf{Avg.\ reward} \\
        \midrule
        \textbf{HRPO (full)} & \textbf{15.33} & \textbf{0.6869} & \textbf{10.5280} & \textbf{0.5480} \\
        w/o cohorting & 14.72 & 0.6338 & 9.3310 & 0.4280 \\
        w/o residual credit & 15.13 & 0.6696 & 10.1320 & 0.4880 \\
        w/o credit-to-go & 15.14 & 0.6522 & 9.8720 & 0.4670 \\
        w/o KL & 15.10 & 0.6657 & 10.0490 & 0.4930 \\
        \bottomrule
    
      \end{tabular}}
    
    \end{table}

\begin{figure*}[t]
  \centering
  \includegraphics[width=0.70\textwidth]{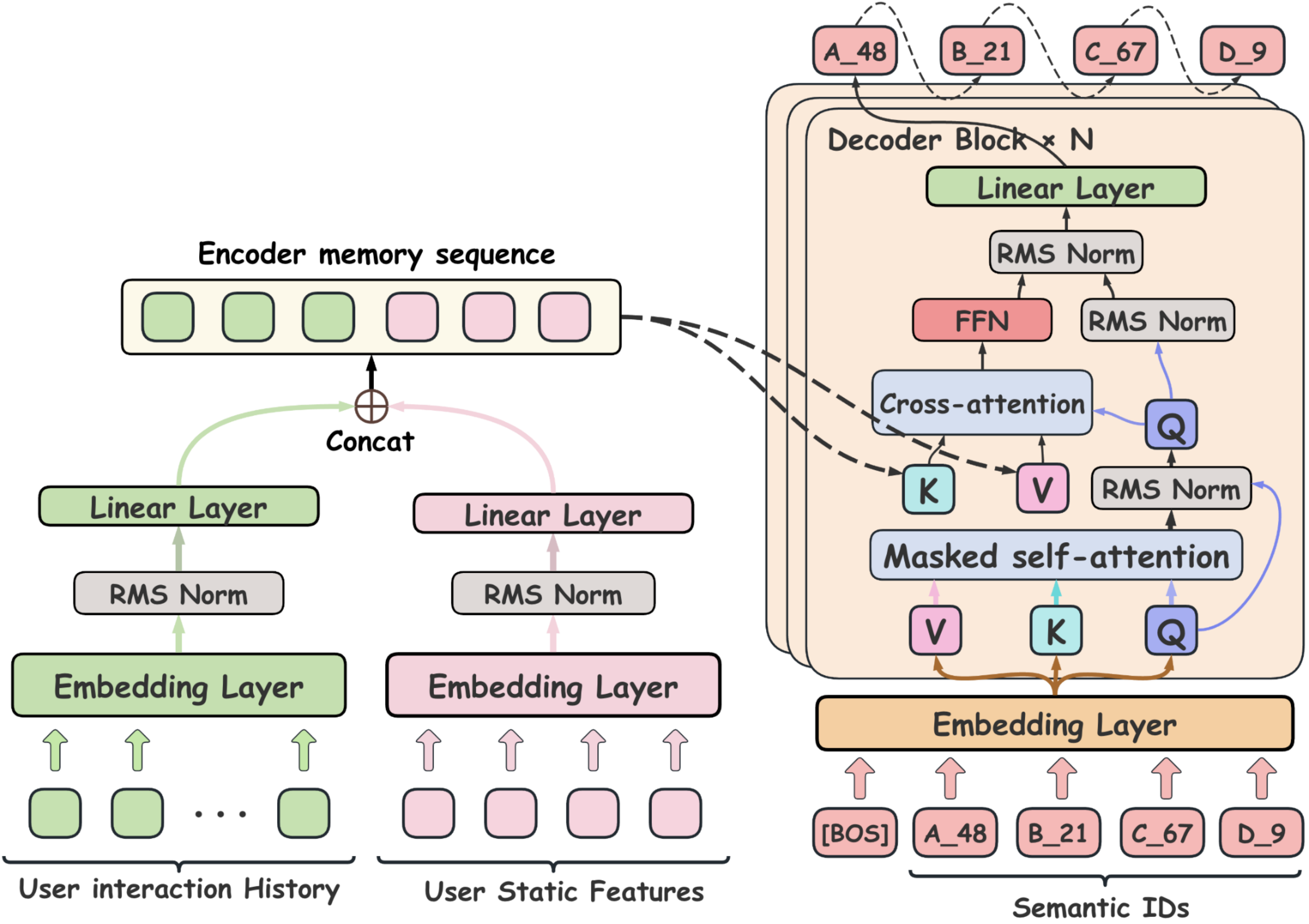}
  \vspace{-1.0em}
  \caption{Production-backbone architecture used for A/B serving and the scaling-law diagnostic.}
  \label{fig:business_architecture}
\end{figure*}

\subsection{Component Analysis (RQ3)}
\label{sec:exp_component}

Table~\ref{tab:ablation_hrpo_AB_compact} studies four ablations of \textsc{HRPO} by removing one component at a time: cohorting, residual credit, credit-to-go, and KL regularization.
The results reveal a clear division of labor across the pipeline.
Cohorting has the largest effect on signal quality: removing it reduces session return from 10.528 to 9.331 and one-step reward from 0.548 to 0.428.
This indicates that stable prefix-utility estimation is the main bottleneck under exposure-limited logs, where raw prefix feedback can be sparse and heterogeneous.

(1) Removing cohorting yields the largest drop in both one-step and whole-session metrics, highlighting the importance of cohort-conditioned prefix utilities for stabilizing group-based updates.
Without cohorting, prefix utilities become less calibrated across heterogeneous contexts, making within-group comparisons noisier and amplifying spurious advantages early in the SID path.

(2) Removing residual credit reduces session return from 10.528 to 10.132 and one-step reward from 0.548 to 0.488.
This supports the need for token-aligned residual advantages that disentangle per-layer contributions along the SID path.
Such localization prevents late-token noise (fine-level errors) from being incorrectly attributed to early coarse decisions, which is crucial for coarse-to-fine decoding.

(3) Removing credit-to-go further reduces session return to 9.872 and one-step reward to 0.467.
This shows that accumulating residual credits over the remaining suffix provides a more informative and lower-variance training signal than using only local token credit.
In particular, credit-to-go makes advantages more comparable at the same SID position by accounting for remaining utility beyond the current prefix baseline, reducing myopic updates.

(4) Finally, removing KL regularization causes a consistent degradation, reducing session return to 10.049 and one-step reward to 0.493.
This suggests that anchoring updates to the SFT initialization is beneficial when optimizing against hard negatives in group-based objectives.
KL acts as a trust region that limits overreaction to a finite proposal set, which otherwise can overfit rare tokens/prefixes and hurt constrained decoding stability.

\subsection{Online Deployment (RQ4)}
\label{sec:exp_online}

\begin{figure}[t]
  \centering
  \includegraphics[width=0.75\linewidth]{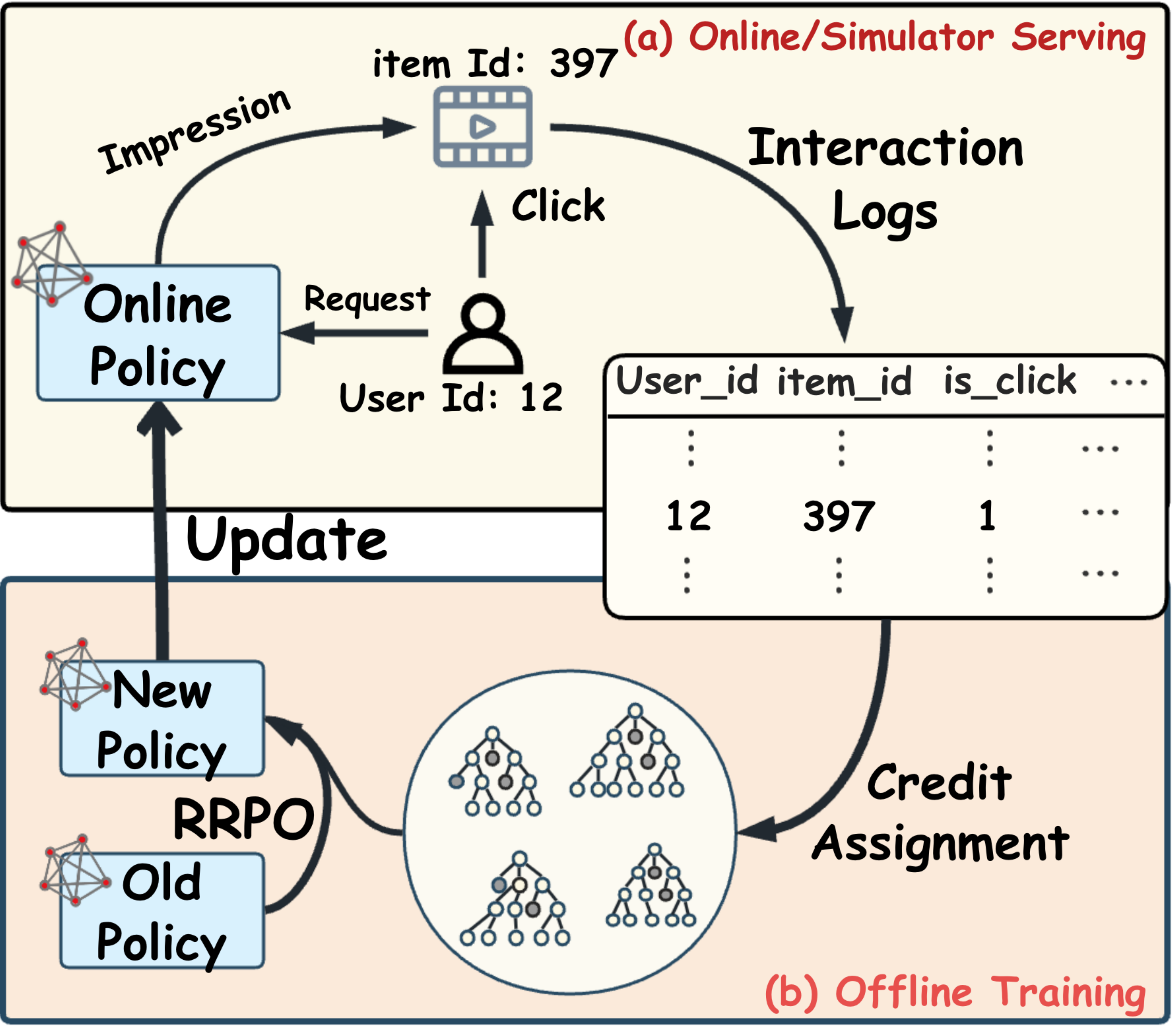}
  \vspace{-0.5em}
  \caption{Online/simulator serving and offline RRPO training loop.}
  \label{fig:ab}
\end{figure}

We validate whether the offline improvements translate to a real production setting with an online A/B experiment in a large-scale advertising system.
Figure~\ref{fig:business_architecture} shows the production backbone used for A/B serving.
The encoder memory sequence is formed by concatenating interaction-history tokens and static-feature tokens, and the SID decoder uses cross-attention to this memory while autoregressively decoding SID tokens.
Figure~\ref{fig:ab} summarizes the deployment loop shared by online/simulator serving and RRPO-style continuation.
In panel (a), the deployed SID policy serves requests and generates impressions; user responses are recorded and stored as interaction logs.
In panel (b), we convert logged outcomes into token-level credits via hierarchical credit assignment and optimize the SID decoder with RRPO to obtain an updated policy, which is periodically deployed to refresh the online policy.
Here, the old policy denotes the frozen reference policy used to generate proposal groups and compute PPO-style likelihood ratios during RRPO training, rather than an estimate of the historical online behavior policy.
Thus, RRPO does not require logged serving probabilities; instead, the old and updated policies are evaluated on the same SID prefixes, and the frozen reference serves as a conservative anchor for token-level updates.

Table~\ref{tab:online_ab_segment} reports the relative changes of \textsc{HRPO} versus control on three target IAA traffic segments.
Exposure measures served traffic volume and is used as a guardrail.
Cost measures observed business cost, while Target Cost is the calibrated target-cost metric used in the launch review.
The results show positive Target Cost lifts across all three segments, with the strongest gain on Fiction IAA, while Exposure and Cost remain close to the control policy.
This pattern suggests that the online gain is not driven by a large traffic-volume shift or by uniformly increasing cost.
Instead, \textsc{HRPO} improves the target metric within the existing serving distribution, which is consistent with its conservative token-level update design and the deployment requirement of preserving stable business traffic.

\begin{table}[t]
\centering
\caption{Online A/B results by in-app advertising (IAA) traffic segment.}
\label{tab:online_ab_segment}
\vspace{-1.0em}
\footnotesize
\setlength{\tabcolsep}{4.0pt}
\renewcommand{\arraystretch}{1.05}
\begin{tabular*}{\linewidth}{@{\extracolsep{\fill}}lrrr@{}}
\toprule
Segment & Exposure & Cost & Target Cost \\
\midrule
Short-Drama IAA & -0.117\% & +0.024\% & +0.168\% \\
Mini-Game IAA & -0.132\% & -0.283\% & +0.186\% \\
Fiction IAA & +0.750\% & +0.823\% & +3.490\% \\
\bottomrule
\vspace{-3.0em}
\end{tabular*}

\end{table}

\section{Related Work}

\subsection{Generative retrieval and semantic identifiers}
Large-scale retrieval and recommendation search enormous action spaces under efficiency constraints~\cite{covington2016youtube,embeddingSurvey}.
Score-based sequential recommendation predicts/ranks the next item from history (e.g., GRU4Rec~\cite{gru4rec}, SASRec~\cite{sasrec}, and LLM-based sequential models~\cite{llmSeqRecArch}), while structured retrieval exploits hierarchical indexing (e.g., TDM~\cite{tdm}) to scale search.
Generative retrieval replaces nearest-neighbor search with constrained identifier decoding~\cite{dsi}.

Building on structured identifiers, SID-based generative recommenders (e.g., TIGER~\cite{tiger} and follow-ups~\cite{onerec,minionerec,gflowgr,nezha}) scale recommendation via validity-constrained decoding.
Recent tokenization studies improve action/item and user-context representations: LETTER learns item tokenizers, ActionPiece tokenizes action sequences with context, Pctx tokenizes personalized context, and MTGRec pre-trains with multi-identifier item tokenization~\cite{letter,actionpiece,pctx,mtgrec}.
However, they mainly use next-token imitation~\cite{tiger,onerec,minionerec}, and post-training often remains item/sequence-level~\cite{dpo,sdpo,sprec,rere}.
This leaves a log-only gap: turning exposure-limited item feedback into token-aligned, layer-sensitive signals for stable token-wise updates~\cite{bottou-jmlr-2013,li2011unbiased,swaminathan2015crm,dudik2011doublyrobust}.
We address this by token-wise credit design for SID decoding, complementing item-level post-training~\cite{sdpo,sprec,rere}.

\subsection{Preference Optimization for Generative Recommendation}
Preference-based post-training aligns generative models beyond imitation via RLHF-style PPO updates~\cite{christiano2017preferences,ouyang2022instructgpt,ppo}.
Preference-only objectives such as DPO optimize a supervised log-ratio without reward modeling~\cite{dpo}, and group-based variants improve stability via within-group normalization~\cite{shao2024deepseekmath}.

In recommendation, S-DPO adapts DPO to list-wise learning with Plackett--Luce objectives and multiple negatives~\cite{sdpo,plackett1975analysis}, while SPRec uses self-play style data construction for LM-based recommenders~\cite{sprec,geng2022p5,jiang-etal-2025-reclm}.
RL-based post-training has also been explored, e.g., ReRe applies RL-driven reranking over generated candidates~\cite{rere,gflowgr}.
Yet most methods optimize at the item/list level, providing coarse supervision for internal token decisions of hierarchical identifier generators.
We instead derive structured token-wise credits from logged outcomes and optimize the SID decoder with conservative, PPO-style token-level updates~\cite{ppo}.

\subsection{Interactive Policy Learning for Recommendation}
Recommendation is sequential with delayed, multi-objective feedback.
Sequential recommenders (GRU4Rec~\cite{gru4rec}, NARM~\cite{narm}, SASRec~\cite{sasrec}, BERT4Rec~\cite{bert4rec}) target next-item prediction, while RL formulations optimize session-level objectives~\cite{pagewiseRec}; for slate actions, SlateQ decomposes slate value for tractable learning~\cite{ijcai2019p360}.

Prior work studies off-policy RL (DDPG~\cite{ddpg}, TD3~\cite{td3}), pairwise learning from negative feedback~\cite{negativeFeedbackPDRL}, latent-action exploration and regularization~\cite{latentActionSpace}, and hierarchical RL (HAC~\cite{hac}) for delayed feedback and competing objectives~\cite{zou2019longterm,generativeAutoBidding}.
Because online iteration is costly, simulators support controlled evaluation: RecSim~\cite{recsim} is a general framework, and KuaiSim~\cite{kuaisim} is an industrial-scale simulator benchmarked on logs such as KuaiRand~\cite{kuairand}.
Reward shaping motivates densifying sparse rewards under suitable conditions~\cite{ng1999shaping}.

We use simulators for policy evaluation and continued improvement.
Our PPO-compatible, token-level objective enables a hierarchical semantic-ID generator to start from logs and seamlessly fine-tune further with rollouts, motivating comparisons to RL baselines while we focus on stable optimization under sparse multi-behavior feedback via reward densification and cross-layer token credit assignment.
Thus, HRPO bridges contextual bandit-style logged learning and simulator-interactive RL by keeping the same token-level update rule.

\section{Conclusion}

In this work, we present \textsc{HRPO}, a post-training framework for SID-based generative recommendation under exposure-limited offline logs with only terminal, item-level feedback. \textsc{HRPO} estimates reliable utilities for SID prefixes and decomposes them into residual, token-aligned credits, which are optimized via a conservative RRPO objective for stable offline improvement. Experiments show that \textsc{HRPO} delivers substantial gains in simulator rollouts and transfers to production, improving Target Cost across target IAA traffic segments while keeping exposure and cost close to the control policy in an online A/B experiment. Future work will extend residual credit assignment beyond trie-structured SIDs and improve robustness under distribution shift and long-tail sparsity in broader industrial serving settings.

\begin{acks}
This research was partially supported by National Natural Science Foundation of China (No. 62502404), Hong Kong Research Grants Council (Research Impact Fund No. R1015-23, Collaborative Research Fund No. C1043-24GF, General Research Fund No. 11218325), Institute of Digital Medicine of City University of Hong Kong (No. 9229503), and Kuaishou (CCF-Kuaishou Large Model Explorer Fund No. 2025008, Kuaishou University Cooperation Project).
\end{acks}

\bibliographystyle{ACM-Reference-Format}
\balance
\bibliography{9Reference}

\appendix

\section{HRPO Algorithms}
\label{app:hrpo_algorithms}

This appendix summarizes the offline post-training loop used in the paper.
The procedure optimizes the RRPO objective (Eq.~\eqref{eq:rrpo_obj}) using
cohort-conditioned smoothed prefix means (Eq.~\eqref{eq:shrink}) and residual credit-to-go
(Eqs.~\eqref{eq:residual_return_def}--\eqref{eq:residual_return_telescoping}).

\begin{algorithm}[H]
\caption{HRPO Post-Training (Offline)}
\label{alg:hrpo_offline}
\begin{algorithmic}[1]
\Require Logged dataset $\mathcal{D}=\{(u_i,x_i,v_i,\mathbf{b}_i)\}$; mapping $\phi(\cdot)$; admissible set $\mathcal{A}(\cdot)$.
\Require Pretrained policy $\pi_\theta$; frozen reference policy $\pi_{\theta_{\mathrm{old}}}$.
\Require Max cohorts $C$; group size $W$; smoothing $\alpha$; clip $\epsilon_{\text{clip}}$; $\beta_{\mathrm{KL}}$; small constant $\varepsilon$; sync period $M_{\text{sync}}$.

\vspace{0.25em}
\State (Cohorts) Build cohort ids $c(u)$ and include a global cohort $c=-1$ (Eq.~\eqref{eq:cohort_def}).
\State (Prefix tables) Initialize prefix stats $n(c,y_{1:t}),s(c,y_{1:t})$ for $c\in\{-1,0,\dots,C-1\}$ and all trie prefixes $y_{1:t}$.
\For{each $(u,x,v,\mathbf{b})\in\mathcal{D}$}
  \State $y\!\leftarrow\!\phi(v)$,\; $r\!\leftarrow\!r(\mathbf{b})$ (Eq.~\eqref{eq:scalar_utility}); update $(n,s)$ for all prefixes of $y$ for both $c(u)$ and $c=-1$ (Eq.~\eqref{eq:prefix_stats}).
\EndFor
\State Compute smoothed prefix means $\hat\mu(c,y_{1:t})$ (Eq.~\eqref{eq:shrink}).

\vspace{0.25em}
\For{each training step $s=1,2,\dots$}
  \State Sample a minibatch $\{(u_i,x_i,v_i,\mathbf{b}_i)\}_{i=1}^{B}$ from $\mathcal{D}$.
  \For{each instance $(u_i,x_i,v_i,\mathbf{b}_i)$ in the minibatch}
    \State $c_i\leftarrow c(u_i)$;\;\; $y_i^{(1)}\leftarrow \phi(v_i)$.
    \State (Proposals) $Y(x_i)\leftarrow\{y_i^{(1)}\}\cup$ up to $W\!-\!1$ additional unique candidates via constrained decoding under $\pi_{\theta_{\mathrm{old}}}$ with $y_t\in\mathcal{A}(y_{<t})$, then deduplicate.
    \State (Returns \& advantages) For each $y_i^{(w)}\in Y(x_i)$, compute $\{\hat G_t^{(w)}\}_{t=1}^L$ from $\hat\mu(c_i,\cdot)$ by Eq.~\eqref{eq:residual_return_telescoping}, then normalize across $w$ per position to obtain $\{A_t^{(w)}\}$ (Eq.~\eqref{eq:token_group_adv}).
  \EndFor
  \State (RRPO update) Update $\theta$ by maximizing Eq.~\eqref{eq:rrpo_obj} on the minibatch with trust-region penalty $\mathrm{KL}_{\mathrm{TR}}$ (Eq.~\eqref{eq:tr_kl}).
  \If{$s \bmod M_{\text{sync}} = 0$}
    \State Sync reference: $\theta_{\mathrm{old}} \leftarrow \theta$.
  \EndIf
\EndFor
\end{algorithmic}
\end{algorithm}

\subsection{Offline HRPO Post-Training}
\label{app:hrpo_offline_algo}

Offline training first precomputes cohort-conditioned prefix statistics from $\mathcal{D}$ and then repeats:
(i) proposal generation under the frozen reference policy $\pi_{\theta_{\text{old}}}$,
(ii) token-wise advantages from prefix tables via residual credit-to-go, and
(iii) RRPO updates with periodic reference-policy synchronization (Alg.~\ref{alg:hrpo_offline}).
This procedure yields a stable post-trained decoder while keeping updates aligned with the logged-data support.

\subsection{Cohort Construction and Sensitivity}
\label{app:cohort_sensitivity}

For the default cohort mapping $g(u)$ in Eq.~\eqref{eq:cohort_def}, we use deterministic tuple bucketing over stable, coarse-grained user-side attributes available in KuaiRand. The mapping is fixed before post-training, ignores empty buckets, and yields 120 effective cohorts in the default setting. Tables~\ref{tab:grouping_sensitivity}--\ref{tab:cohort_support} report grouping sensitivity and cohort-support statistics.
These results show a support--granularity trade-off. Without cohorting, global prefix statistics are too coarse for heterogeneous users. Coarse buckets under-partition users, whereas overfine buckets fragment support and make prefix estimates noisy. Default buckets and KMeans clusters remain close, suggesting that HRPO benefits from cohort-level smoothing rather than from a specific handcrafted partition.

\begin{table}[t]
\centering
\caption{Grouping sensitivity under different cohort construction strategies.}
\label{tab:grouping_sensitivity}
\vspace{-0.8em}
\scriptsize
\setlength{\tabcolsep}{2.0pt}
\renewcommand{\arraystretch}{1.03}
\resizebox{\linewidth}{!}{%
\begin{tabular}{l c c c c}
\toprule
Strategy & Session length & Avg. reward & Total reward & One-step reward \\
\midrule
No cohorting & 14.72 & 0.6338 & 9.3310 & 0.4280 \\
Coarse buckets & 15.01 & 0.6615 & 9.9300 & 0.4960 \\
Default buckets & 15.33 & 0.6869 & 10.5280 & 0.5480 \\
KMeans clusters & 15.27 & 0.6812 & 10.4010 & 0.5330 \\
Overfine buckets & 15.09 & 0.6704 & 10.1180 & 0.5070 \\
\bottomrule
\end{tabular}}
\vspace{-0.4em}
\end{table}

\begin{table}[t]
\centering
\caption{Cohort support statistics for non-empty cohort construction strategies.}
\label{tab:cohort_support}
\vspace{-0.8em}
\scriptsize
\setlength{\tabcolsep}{2.0pt}
\renewcommand{\arraystretch}{1.03}
\resizebox{\linewidth}{!}{%
\begin{tabular}{l c c c c c}
\toprule
Strategy & Effective cohorts & P25 & Median & P75 & Mean \\
\midrule
Coarse buckets & 63 & 87 & 133 & 367 & 380.14 \\
Default buckets & 120 & 34 & 62 & 137 & 216.34 \\
KMeans clusters & 192 & 12 & 42 & 84 & 142.11 \\
Overfine buckets & 424 & 1 & 5 & 24 & 64.35 \\
\bottomrule
\end{tabular}}
\vspace{-0.4em}
\end{table}

\section{Evaluation on Agent4Rec-MovieLens}
\label{app:agent4rec_movielens}

To further examine whether the benefit of \textsc{HRPO} transfers beyond the KuaiRand/KuaiSim environment, we conduct an additional evaluation on the public Agent4Rec-MovieLens simulator~\cite{zhang2024agent4rec}.
Agent4Rec constructs LLM-powered user agents from MovieLens-1M profiles and simulates page-by-page recommendation sessions.
We evaluate all methods under the same simulator setup and report two session-level metrics: AvgR, the average reward over simulated interactions, and Avg. Session Length, the average number of recommendation pages consumed before the user leaves.

Table~\ref{tab:agent4rec_movielens} shows that TIGER+\textsc{HRPO} achieves the best performance on both metrics.
Compared with TIGER, \textsc{HRPO} improves AvgR from 0.853 to 0.884 and session length from 8.33 to 9.12, also outperforming DPO and GRPO on the same backbone. Since Agent4Rec differs from KuaiSim in simulator construction and user modeling, this result suggests the gain is not tied to one rollout environment.

\begin{table}[H]
\centering
\caption{Evaluation on the public Agent4Rec-MovieLens simulator. Higher is better for both metrics.}
\label{tab:agent4rec_movielens}
\vspace{-0.4em}
\small
\setlength{\tabcolsep}{6pt}
\renewcommand{\arraystretch}{1.08}
\begin{tabular*}{0.85\linewidth}{@{\extracolsep{\fill}}lcc@{}}
\toprule
Model & AvgR & Avg. Session Length \\
\midrule
Random & 0.558 & 4.67 \\
GRU4Rec & 0.797 & 6.96 \\
SASRec & 0.859 & 7.79 \\
TD & \underline{0.878} & 8.85 \\
DDPG & \underline{0.878} & \underline{9.02} \\
HAC & 0.873 & 8.94 \\
A2C & 0.869 & 8.73 \\
TIGER & 0.853 & 8.33 \\
TIGER+DPO & 0.861 & 8.48 \\
TIGER+GRPO & 0.869 & 8.70 \\
TIGER+\textsc{HRPO} & \textbf{0.884} & \textbf{9.12} \\
\bottomrule
\end{tabular*}
\vspace{-0.4em}
\end{table}

\section{Scaling-Law Logs for A/B Models}
\label{app:scaling_diagnostics}

In addition to the main HRPO experiments, we report auxiliary scaling-law diagnostics
from six A/B model sizes used in production training logs.

\begin{figure}[H]
  \centering
  \includegraphics[width=0.79\linewidth]{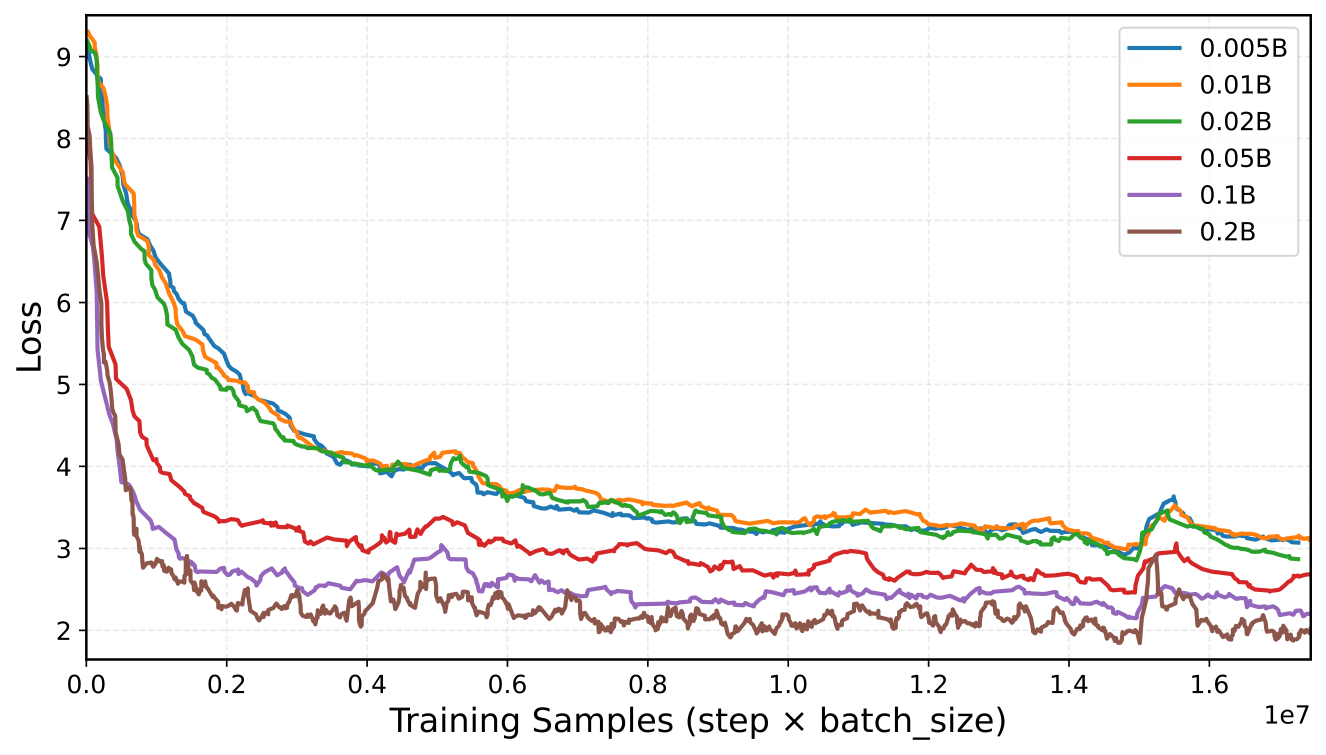}
  \vspace{-1.0em}
  \caption{Scaling-law diagnostic for A/B model sizes.}
  \label{fig:scaling_trunc}
\end{figure}

Table~\ref{tab:iaa_model_sizes} summarizes the model-size and training settings used in the scaling study.
Here, Size denotes the number of model parameters (in billions, e.g., 0.2B = 200M parameters).
Batch size is chosen per model scale based on our production training recipe; learning rate is tuned per scale accordingly.
Figure~\ref{fig:scaling_trunc} plots the corresponding training loss curves. Larger models generally reach lower loss faster, and all curves decrease smoothly without unstable spikes. We include these logs as a diagnostic complement to the online experiment, showing that production training remains numerically stable across the A/B model-size range, while the main HRPO conclusions still come from the controlled offline, simulator, and online comparisons.

\begin{table}[H]
\caption{A/B model configurations for the scaling-law diagnostic.
FLOPs are reported with batch size 1.
Abbreviations: $d$=$d_{\text{model}}$ (hidden width), \#L=\#layers, \#H=\#heads, emb=embedding dimension,
bs=batch size, lr=learning rate.}
\label{tab:iaa_model_sizes}
\centering
\scriptsize
\setlength{\tabcolsep}{5.0pt}
\renewcommand{\arraystretch}{1.05}
\resizebox{0.95\linewidth}{!}{%
\begin{tabular}{l r r r r r r r}
\toprule
Size & FLOPs & $d$ & \#L & \#H & emb & bs & lr \\
\midrule
0.005B & 20M    & 64   & 6  & 4  & 16 & 896 & 0.0030 \\
0.01B  & 40.6M  & 128  & 8  & 4  & 16 & 512 & 0.0010 \\
0.02B  & 83.2M  & 256  & 12 & 8  & 16 & 256 & 0.0005 \\
0.05B  & 160.5M & 512  & 6  & 8  & 16 & 128 & 0.0004 \\
0.1B   & 243.8M & 768  & 8  & 12 & 16 & 64  & 0.0003 \\
0.2B   & 332.7M & 1024 & 12 & 16 & 16 & 32  & 0.0001 \\
\bottomrule
\end{tabular}}
\end{table}

\end{document}